\documentstyle[aps,multicol,epsfig]{revtex}

\newcommand{\simle}
{\raisebox{-0.75ex}[-1.5ex]{$\;\stackrel{<}{\sim}\;$}}

\def\d{\partial}
\def\s{{\sigma}}
\def\e{{\epsilon}}
\def\k{{ {\bf k} }}

\def\q{{ {\bf q} }}
\def\Q{{ {\bf Q} }}

\def\w{{\omega}}

\def\g{{\gamma}}
\def\i{{ {\rm i} }}

\begin{document}

\def\runtitle{
Magnetoresistance in High-$T_{\rm c}$ Superconductors:
The Role of Vertex Corrections
}
\def\runauthor
 {Hiroshi {\sc Kontani}}

\title{
Magnetoresistance in High-$T_{\rm c}$ Superconductors: \\
The Role of Vertex Corrections
}

\author{
Hiroshi {\sc Kontani}$^{1,2}$
}

\address{
$^1$Theoretische Physik III, Elektronische Korrelationen und Magnetismus, 
Universit\"at Augsburg, D-86135 Augsburg, Germany
\\
$^2$Department of Physics, Saitama University,
255 Shimo-Okubo, Urawa-city, 338-8570, Japan
}

\date{\today}
  
\maketitle

\begin{abstract}
In high-$T_{\rm c}$ cuprates,
the orbital magnetoresistance in plane (${\mit\Delta}\rho/\rho$)
is anomalously enhanced at lower tempemeratures
compared with conventional Fermi liquids, 
and thus Kohler's rule is strongly violated.
Moreover, it should be noted that an intimate relation between 
${\mit\Delta}\rho/\rho$ and the Hall coefficient ($R_{\rm H}$),
$\Delta\rho/\rho \propto (R_{\rm H}/\rho)^{2}$,
holds well experimentally, and is called the "modified Kohler's rule".
In this letter,
we study this long-standing problem
in terms of the nearly antiferromagnetic (AF) Fermi liquid.
We analyze the exact expression for ${\mit\Delta}\rho/\rho$
by including the vertex corrections (VC's) to keep
the conservation laws, and find the approximate "scaling relation"
${\mit\Delta}\rho/\rho \propto \xi_{\rm AF}^4 \cdot \rho^{-2} \cdot H^2$
($\xi_{\rm AF}$ being the AF correlation length.)
in the presence of AF fluctuations.
The factor $\xi_{\rm AF}^4$, which comes from 
the VC's for the current,
gives the additional temperature dependence.
By taking account of the relation
$R_{\rm H} \propto \xi_{\rm AF}^2$
[Kontani et al., PRB 59 (1999) 14723.],
we can naturally explain the modified Kohler's rule.
In conclusion,
based on the Fermi liquid theory,
the famous {\it seemingly} non-Fermi liquid behaviors of
$R_{\rm H}$ and ${\mit\Delta}\rho/\rho$
in high-$T_{\rm c}$ cuprates are naturally understood 
on an equal footing.
\end{abstract}


\sloppy

\begin{multicols}{2}

Anomalous normal-state transport properties 
in high-$T_{\rm c}$ superconductors
have attracted a great deal of attentions.
For a wide range of temperatures,
the resistivity ($\rho$),
the Hall coefficient ($R_{\rm H}$), and
the in-plane magnetoresistance (MR, ${\mit\Delta}\rho/\rho$)
are approximately
proportional to $T$, $T^{-1}$ and $T^{-4}$,
respectively
 \cite{Harris,Kimura,Tyler,Ando}.
These findings indicate the strong violation of Kohler's rule, that is,
$R_{\rm H}\propto \rho^0$ and ${\mit\Delta}\rho/\rho\propto \rho^{-2}$.
Kohler's rule is derived by the relaxation time approximation
(RTA), assuming that the relaxation time ($\tau_\k$) is not very
anisotropic.
Irrespective of its quasiclassical approximation,
Kohler's rule holds in various metals in general
 \cite{Ziman}.

By definition,
$\rho$, $R_{\rm H}$, and
${\mit\Delta}\rho/\rho$ are given by
\begin{eqnarray}
\rho &=& 1/\s_{xx} \nonumber \\
R_{\rm H} &=& ({\mit\Delta}\s_{xy}/H)/\s_{xx}^2,
 \label{eqn:coeff-def} \\
{\mit\Delta}\rho/\rho &=& -{\mit\Delta}\s_{xx}/\s_{xx} 
 - ({\mit\Delta}\s_{xy}/\s_{xx})^2,
 \nonumber 
\end{eqnarray}
where $\s_{xx}$ is the diagonal conductivity without 
the magnetic field ${\vec H}$ (${\vec H}\parallel{\hat e}_z$).
The Hall conductivity ${\mit\Delta}\s_{xy}$ and 
the magnetoconductivity (MC) ${\mit\Delta}\s_{xx}$ are
of the order $H^1$ and $H^2$, respectively.

The violation of Kohler's rule has been regarded as 
a serious test for the possible ground state of 
high-$T_{\rm c}$ cuprates.
Until today,
various authors have attempted to settle this issue
in terms of a non-Fermi liquid picture, e.g.,
the spin-charge separation picture \cite{Anderson}.
In recent years,
several authors studied this problem 
in terms of the nearly antiferromagnetic (AF) 
Fermi liquid picture, and 
they ascribed the violation of Kohler's rule 
to the anisotropy of $\tau_\k$ based on the RTA
 \cite{Pines,Millis}.
However, they had to assume a huge anisotropy of $\tau_\k$
to produce the magnitude of the Hall coefficient.
In that case, their model produced a MR
too large to agree with experiments.

Recently,
we studied the role of vertex corrections (VC's)
on the Hall coefficient in high-$T_{\rm c}$ cuprates
 \cite{Kontani,Kanki}.
We found that the VC's for the current,
which are dropped in the RTA,
cause the Curie-Weiss-like behavior of $R_{\rm H}$ 
in the presence of AF fluctuations.
The obtained results explain the experimental doping and
temperature dependences of $R_{\rm H}$ satisfactorily
 \cite{Sato}.
Noteworthily,
$R_{\rm H}<0$ observed in electron-doped compounds,
which cannot be explained within the analysis by the RTA,
is naturally reproduced in terms of the VC's
 \cite{Kontani,Kanki}.

In this letter,
we introduce our theoretical study on the MR 
in high-$T_{\rm c}$ cuprates
according to the expressions for the MR
which is "exact" with respect to $O(\tau^2)$
 \cite{MR-formula}.
We take account of all the VC's required by the Ward identity
in order not to violate the conservation laws
(i.e., the conserving approximation)
 \cite{Baym}.
Due to the VC's, the MR is strongly enhanced
in the presence of AF fluctuations,
which leads to the violation of Kohler's rule
in high-$T_{\rm c}$ cuprates.
Furthermore, the so-called "modified Kohler's rule",
${\mit\Delta}\rho/\rho \propto (R_{\rm H}/\rho)^{2}$,
is well reproduced in our study in terms of the 
nearly AF Fermi liquid.

First,
we calculate the self-energy $\Sigma_\k(\w)$
by the fluctuation-exchange (FLEX) approximation,
which is a kind of self-consistent perturbation theory with respect to $U$
 \cite{Bickers}.
It is suitable for the analysis of the nearly AF Fermi liquid,
and it has been applied to high-$T_{\rm c}$ cuprates
 \cite{Dahm}
and $\kappa$-BEDT-TTF superconductors
 \cite{Kino,Kondo,Schmalian}.
In the present study, we use $64\times64$ $\k$-meshes and
512 Matsubara frequencies.

\begin{figure}
\begin{center}
\epsfig{file=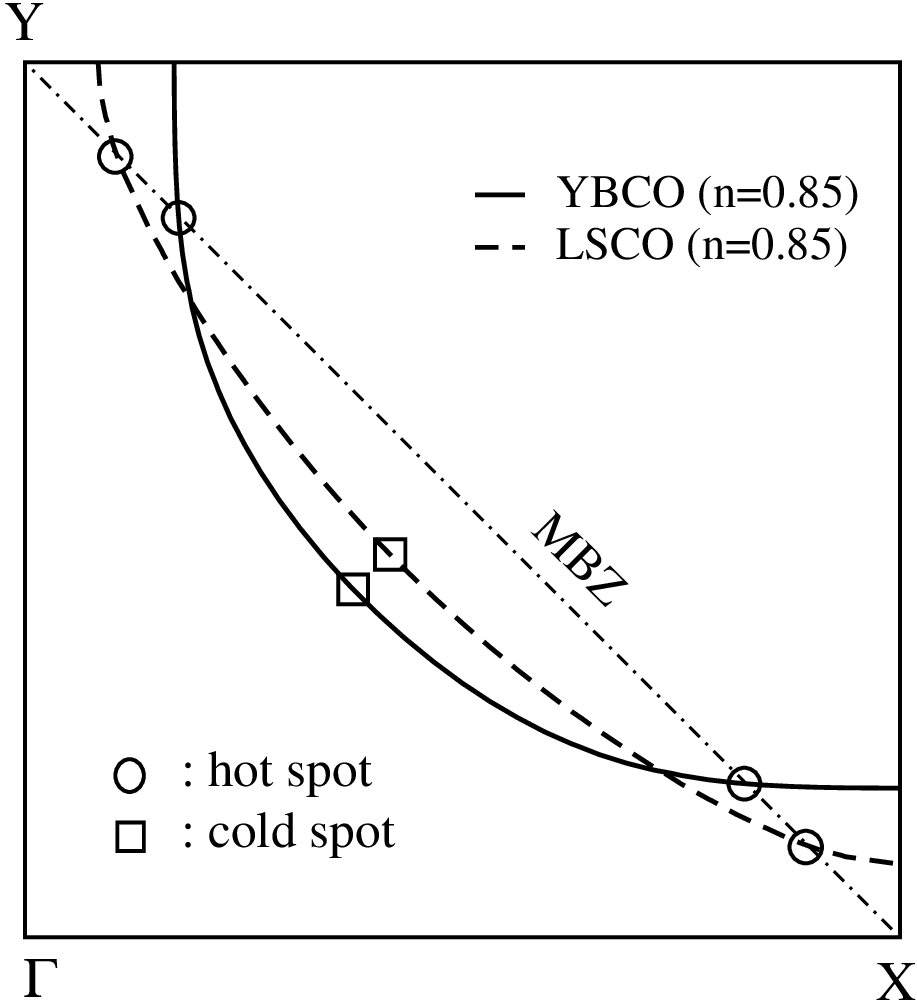,width=5cm}
\end{center}
\caption{}
{\small 
The noninteracting FS for
YBCO and for LSCO.
On the FS, $\g_\k$ takes the maximum [minimum] value
around the hot-spot [cold-spot]
(see ref.~\cite{Kontani}).
}
  \label{fig:FS85-free}
\end{figure}

We study the square-lattice Hubbard model with on-site 
Coulomb interaction $U$.
Figure~\ref{fig:FS85-free} shows the noninteracting 
Fermi surface (FS) for $n=0.85$.
We put $(t,t',t'';U)=(1,-0.2,0.15;6.5)$ for YBCO, and
$(1,-0.15,0.05; 4.5)$ for LSCO,
where $t$, $t'$ and $t''$ are hopping integrals
between the nearest neighbor, the second nearest and the third 
nearest, respectively
 \cite{Comment}.
As shown in ref.~\cite{Kontani},
$\g_\k\equiv{\rm Im}\Sigma_\k(-\i0)>0$ takes a 
maximum [minimum] value 
around the hot-spot [cold-spot] in Fig.~\ref{fig:FS85-free}.
The portion of the FS near the cold-spot mainly contributes 
to the transport phenomena.
We note that the anisotropy of $\g_\k$ is not very large
according to the FLEX approximation;
$\g_{\rm hot}/\g_{\rm cold} \simle 3$ for YBCO (n=0.90) at $T=0.02$
 \cite{Kontani}.
This result is consistent with the recent analysis of the $c$-axis MR
 \cite{Schofield},
or with ARPES measurements except in the pseudo spin-gap region.

Next, we study the transport coefficients
by including the VC's for the current
according to the conserving approximation.
The expression for the conductivity without the magnetic field
and the Hall conductivity are given by 
refs.~\cite{Eliashberg} and \cite{Kohno}, respectively:
\begin{eqnarray}
 \s_{xx}
 &=& \frac{e^2}{v_B} \oint_{\rm FS} 
 \frac{dk_\parallel}{|{\vec v}_\k|} v_{\k x}J_{\k x} \frac{1}{\g_\k} ,
 \label{eqn:sigma-lowT}   \\
 {\mit\Delta}\s_{xy}
 &=& -H \cdot \frac{e^3}{4v_B} \oint_{\rm FS} dk_\parallel 
 |{\vec J}_\k|^2 \left({\d\theta_\k^J}/{\d k_\parallel}\right)
  \frac{1}{\g_\k^2} ,
 \label{eqn:sigmaH-lowT}   \\
  J_{\k\mu}
  &=& \frac1{v_B} \oint_{\rm FS}\frac{dk'_\parallel}{|{\vec v}_{\k'}|} 
  {\cal T}(\k|\k') \frac{1}{\g_{\k'}} v_{\k'\mu} 
  + v_{\k\mu} ,
 \label{eqn:J} \\
  & &{\cal T}(\k|\k') \equiv
   \int\frac{d\e}{4\i}{\cal T}_{22}(\k,0|\k',\e) ,
  \label{eqn:T} 
\end{eqnarray}
where $v_B=(2\pi)^2$ and
$\oint_{\rm FS} dk_\parallel$ means the integration on the Fermi line.
${\vec v}_\k \equiv {\vec\nabla}(\e_\k^0+{\rm Re}\Sigma_\k(0))$ 
is the quasiparticle velocity, and 
$\theta_\k^J = {\rm tan}^{-1}(J_{\k y}/J_{\k x})$.
The total current ${\vec J}_\k$ is
given by the solution of the Bethe-Salpeter equation, 
eq.~(\ref{eqn:J}).
Here, the "reducible" vertex ${\cal T}_{22}(\k\e|\k'\e')$ in 
eq.~(\ref{eqn:T}) within the FLEX approximation is explained 
in ref.~\cite{Kontani}.
Note that
${\vec J}_\k$ is not parallel to ${\vec v}_\k$ 
due to ${\cal T}_{22}$ in general.

According to ref.~\cite{MR-formula},
the expression for the MC with the VC's
is given by
\begin{eqnarray}
  {\mit\Delta}\s_{xx}
 &\equiv& {\mit\Delta}\s_{xx}^{a} + {\mit\Delta}\s_{xx}^{b} ,
  \label{eqn:MC-lowT} \\
   {\mit\Delta}\s_{xx}^{a} 
 &=& -H^2 \cdot \frac{e^4}{4v_B} \oint_{\rm FS} 
 \frac{dk_\parallel}{|{\vec v}_\k|} \left\{ d_x(\k) \right\}^2
 \frac{1}{\g_\k},
  \nonumber \\
   {\mit\Delta}\s_{xx}^{b} 
 &=& -H^2 \cdot \frac{e^4}{4v_B} \oint_{\rm FS} 
 \frac{dk_\parallel}{|{\vec v}_\k|} d_x(\k)D_x'(\k)\frac{1}{\g_\k} ,
  \nonumber \\
 d_\mu(\k)
  &=& |{\vec v}_\k| \cdot \frac{\d}{\d k_\parallel}
  \left( {J_{\k\mu}}/{\g_\k} \right) ,
  \nonumber \\
 D_\mu'(\k)
  &=& \frac1{v_B} \oint_{\rm FS} \frac{dk'_\parallel}{|{\vec v}_{\k'}|} 
  {\cal T}(\k|\k') \frac1{\g_{\k'}} d_\mu(\k') .
   \nonumber 
\end{eqnarray}
According to eqs.~(\ref{eqn:sigmaH-lowT}) and (\ref{eqn:MC-lowT}),
${\mit\Delta}\s_{xy}$ and ${\mit\Delta}\s_{xx}$ are
of orders $O(\g^{-2})$ and $O(\g^{-3})$, respectively.
If we drop all the VC's coming from ${\cal T}^I(\k|\k')$
in eqs.~(\ref{eqn:sigma-lowT}), (\ref{eqn:sigmaH-lowT}) 
and (\ref{eqn:MC-lowT}), we obtain the results by the RTA
with $\tau_\k = 1/2\g_\k$.
We write them as
$\s_{xx}^{\rm RTA}$, ${\mit\Delta}\s_{xy}^{\rm RTA}$
and ${\mit\Delta}\s_{xx}^{\rm RTA}$, respectively.

Under the fourfold symmetry around the $z$-axis,
${\mit\Delta}\s_{xx}^{a}$ in eq.~(\ref{eqn:MC-lowT})
is rewritten as
\begin{eqnarray}
  {\mit\Delta}\s_{xx}^{a}
 &=& -H^2 \cdot \frac{e^4}{8v_B} \oint_{\rm FS} dk_\parallel 
 |{\vec v}_\k| 
  \nonumber \\
 & &\times
 \left[ |{\vec l}_\k|^2  
 \left({\d\theta_\k^J}/{\d k_\parallel}\right)^2
 + 
 \left({\d|{\vec l}_\k|}/{\d k_\parallel}\right)^2
 \right] \frac1{\g_\k}
 \label{eqn:MC-lowT2}
\end{eqnarray}
where ${\vec l}_\k = {\vec J}_\k/\g_\k$.
In the same way, 
the MC without the VC's,
${\mit\Delta}\s_{xx}^{\rm RTA}$, is given by eq.~(\ref{eqn:MC-lowT2})
if we replace 
${\vec l}_\k \ \rightarrow 
 \ {\vec l}_\k^v \equiv {\vec v}_\k/\g_\k$
and $\theta_\k^J \ \rightarrow 
 \ \theta_\k^v \equiv \tan^{-1}(v_{\k x}/v_{\k y})$.
We note that eq.~(\ref{eqn:MC-lowT}) is the simplified expression of
the original one given in ref.~\cite{MR-formula},
under the condition $\g_\k^\ast \ll T$.
In the present numerical study,
we use the original expression in ref.~\cite{MR-formula}
to obtain more reliable results.

\begin{figure}
\vspace{10mm}
\begin{center}
\epsfig{file=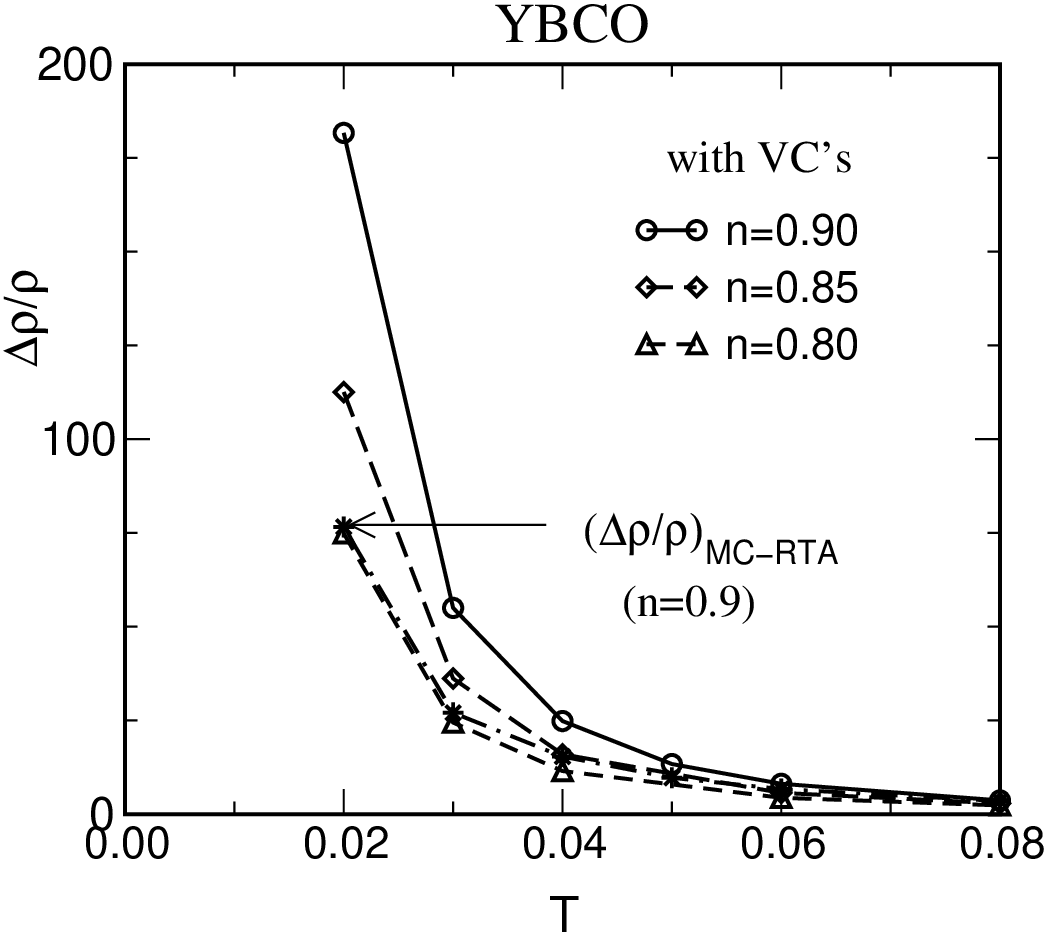,width=6cm}
\end{center}
\vspace{10mm}
\caption{}
{\small 
Temperature dependence of ${\mit\Delta}\rho/\rho$
for YBCO with full VC's.
We see that 
$({\mit\Delta}\rho/\rho)_{\rm MC{\mbox{-}}RTA}$ is smaller,
which means that the VC's for ${\mit\Delta}\s_{xx}$
enhance the MR.
}
  \label{fig:MR-T-YBCO}
\end{figure}

Figure~\ref{fig:MR-T-YBCO}
shows the obtained temperature dependence of ${\mit\Delta}\rho/\rho$
for $T\ge0.02$($\sim$100~K) by using the original expression
of eq.~(\ref{eqn:MC-lowT}).
We also plot $({\mit\Delta}\rho/\rho)_{\rm MC{\mbox{-}}RTA}$,
which is given by replacing ${\mit\Delta}\s_{xx}$
with ${\mit\Delta}\s_{xx}^{\rm RTA}$ in eq.~(\ref{eqn:coeff-def}).
Thus, Fig.~\ref{fig:MR-T-YBCO}
means that the VC's increase the ratio
${\mit\Delta}\s_{xx}/{\mit\Delta}\s_{xx}^{\rm RTA}$
as the temperature decreases.
We will discuss the mechanism later in this letter.

\begin{figure}
\vspace{10mm}
\begin{center}
\epsfig{file=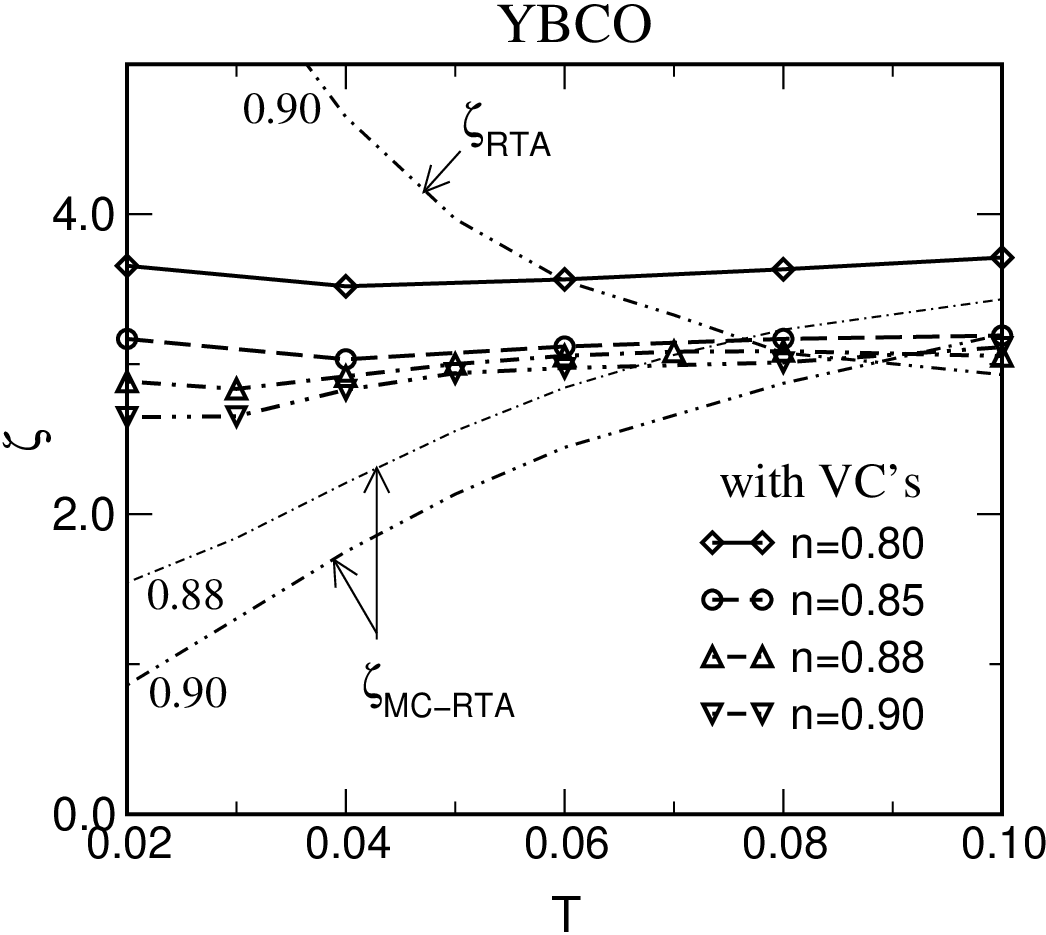,width=7cm}
\end{center}
\vspace{10mm}
\caption{}
{\small 
Calculated $\zeta=\Delta\rho\cdot\rho/R_{\rm H}^2$
for YBCO with full VC's for $0.02\ge T\ge 0.1$
($T=100{\rm K}\sim500$K).
The modified Kohler's rule
$\zeta=$const. holds well only if all the VC's
are taken into account.
} 
  \label{fig:MKR-YBCO}
\end{figure}

Figure~\ref{fig:MKR-YBCO}
shows the temperature dependence of the following coefficient
for YBCO:
\begin{eqnarray}
\zeta&=& ({\mit\Delta}\rho/\rho)\cdot(\cot\theta_{\rm H})^2
 \ = \ 
 -\frac{{\mit\Delta}\s_{xx}\cdot\s_{xx}}{({\mit\Delta}\s_{xy})^2} -1,
   \label{eqn:zeta}
\end{eqnarray}
where $\cot\theta_{\rm H}= \s_{xx}/{\mit\Delta}\s_{xy}
= \rho/R_{\rm H}$.
The calculated $\zeta$ is almost $T$-independent
for a wide range of temperatures, $0.02\le T\le 0.10$,
which is consistent with the "modified Kohler's rule"
observed in high-$T_{\rm c}$ cuprates.
Experimentally,
$\zeta$ slightly depends on the doping;
$\zeta\approx1.5\sim1.7$ for YBCO
 \cite{Harris},
and $\zeta\approx2\sim3$ for Tl$_2$Ba$_2$CuO$_{6+\delta}$
 \cite{Tyler}.
(Note that $\zeta= {\mit\Delta}\rho =0$
for an isotropic system.)

We also show the result of the RTA in Fig.~\ref{fig:MKR-YBCO}, 
$\zeta_{\rm RTA}$,
where all the VC's are neglected.
As was pointed out in previous studies based on the RTA
 \cite{Millis},
$\zeta_{\rm RTA}$ is too large to agree with experiments
at low temperatures.

On the other hand, 
$\zeta_{\rm MC{\mbox{-}}RTA}$ is given by eq.~(\ref{eqn:zeta})
by replacing ${\mit\Delta}\s_{xx}$ with ${\mit\Delta}\s_{xx}^{\rm RTA}$.
Considering that $\s_{xx}\sim\s_{xx}^{\rm RTA}$
as shown in ref.~\cite{Kontani}, we obtain
$\zeta_{\rm MC{\mbox{-}}RTA}+1 \sim (\zeta_{\rm RTA}+1)\cdot
({\mit\Delta}\s_{xy}^{\rm RTA}/{\mit\Delta}\s_{xy})^2$.
As shown in Fig.~\ref{fig:MKR-YBCO},
$\zeta_{\rm MC{\mbox{-}}RTA}$ is too small at low temperatures
because ${\mit\Delta}\s_{xy}$ 
is largely enhanced due to the VC's,
which explains the Curie-Weiss-like behavior of $R_{\rm H}$
 \cite{Kontani}.
In conclusion,
the modified Kohler's rule is well satisfied only when
all the VC's for ${\mit\Delta}\s_{xy}$
and ${\mit\Delta}\s_{xx}$ are taken into account.
Thus, the VC's play an essential role for the MR.
We will discuss the reason for this later.

Here,
we discuss why the modified Kohler's rule 
is well satisfied by taking the VC's into account.
In YBCO, the spin-susceptibility given by the FLEX approximation
is qualitatively described as
\begin{eqnarray}
\chi_\q(\w)= \chi_Q(1+\xi_{\rm AF}^2(\q-\Q)^2-\i\w/\w_{\rm sf})^{-1} ,
 \label{eqn:chi}
\end{eqnarray}
where $\Q=(\pi,\pi)$, $\xi_{\rm AF}$ is the AF-correlation length,
and $\chi_Q \propto \w_{\rm sf}^{-1} \propto \xi_{\rm AF}^2$.
According to the SCR theory \cite{SCR}
or the renormalization group study
 \cite{RG},
$\xi_{\rm AF}^2 \propto (T+c)^{-1}$ in two-dimensional systems,
which is observed in high-$T_{\rm c}$ cuprates around the optimum doping
 \cite{Goto}.
This relation is also reproduced well by the FLEX approximation
 \cite{Dahm}.

As shown in ref.~\cite{Kontani},
${\cal T}^I(\k;\k') \propto
 \i T^2 {\rm Im}\chi_{\k-\k'}(\w)/\w|_{\w\rightarrow0}$
due to the Maki-Thompson term,
which is essential when AF fluctuations are strong.
According to eq.~(\ref{eqn:chi}),
we see that the "momentum-derivative of ${\cal T}_{22}$"
causes a factor proportional to $\xi_{\rm AF}^2 \propto T^{-1}$.
This suggests that 
${\mit\Delta}\s_{xy}$ and ${\mit\Delta}\s_{xx}$
are enhanced by $\xi_{\rm AF}^2$ and $\xi_{\rm AF}^4$ 
according to eqs.~(\ref{eqn:sigmaH-lowT}) and (\ref{eqn:MC-lowT}),
which will cause the breakdown of Kohler's rule
in the presence of AF fluctuations.

Now we discuss ${\mit\Delta}\s_{xx}$ in more detail:
According to the analysis
in \S ~V of ref.~\cite{Kontani},
$\left(\frac{\d\theta_\k^J}{\d k_\parallel}\right) \propto 
\xi_{\rm AF}^{x} \!\cdot\! \left(\frac{\d\theta_\k^v}{\d k_\parallel}\right)$
and $x\approx2$ around the cold spot in YBCO or LSCO.
To be precise,
$x$ may deviate from 2 somehow
due to the $\xi_{\rm AF}$-dependence (or $T$-dependence)
of the shape of the FS, which is prominent
in high-$T_{\rm c}$ cuprates
 \cite{Kontani}.
As a result,
the first term of eq.~(\ref{eqn:MC-lowT2})
is proportional to $\xi_{\rm AF}^{2x}$.

The second term of eq.~(\ref{eqn:MC-lowT2})
vanishes in an isotropic system.
In the case of YBCO,
it is small in our numerical calculation
because $|{\vec v}_\k|$ on the FS is rather isotropic in YBCO.
Although it is enhanced at lower temperatures 
as $\g_\k$ becomes anisotropic,
it is still small in YBCO
because $|{\vec l}_\k| = |{\vec J}_\k/\g_\k|$ is not 
very anisotropic according to the FLEX approximation 
for $n\le0.9$
 \cite{Kontani}.

According to a similar analysis in \S ~V of ref.~\cite{Kontani},
we can show that 
${\mit\Delta}\s_{xx}^{b}$ in eq.~(\ref{eqn:MC-lowT}) 
is also proportional to $\xi_{\rm AF}^{2x}$
 \cite{Future}.
However, its absolute value is quite small in YBCO 
in the present calculation, although this may be accidental.
In conclusion, in the case of YBCO,
only the first term of eq.~(\ref{eqn:MC-lowT2}) is dominant
for ${\mit\Delta}\s_{xx}$
which is enhanced by $\xi_{\rm AF}^{2x}$.

\begin{figure}
\vspace{10mm}
\begin{center}
\epsfig{file=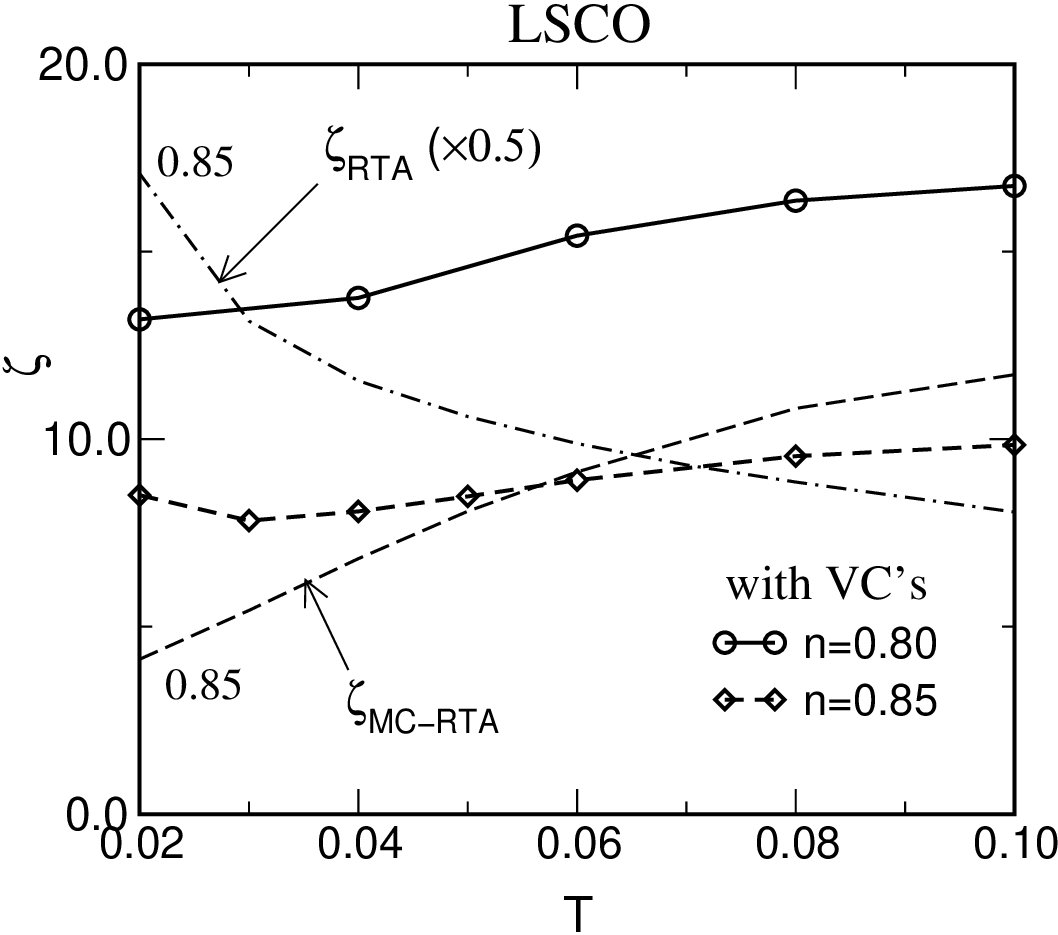,width=7cm}
\end{center}
\vspace{10mm}
\caption{}
{\small 
Temperature dependence of $\zeta$
for LSCO with full VC's.
}
  \label{fig:MKR-LSCO}
\end{figure}

Next, 
we show the temperature dependence of $\zeta$
in the case of LSCO in Fig.~\ref{fig:MKR-LSCO}.
$\zeta$ in LSCO is much larger than that in YBCO,
particularly for $n=0.8$.
This is because
the second term of eq.~(\ref{eqn:MC-lowT2}) becomes important
as the anisotropy of 
$|{\vec v}_\k|$ on the FS is larger in LSCO.
In particular, the FS at $n=0.8$ is very close to
the van-Hove singular point $(\pi,0)$.
Experimentally,
$\zeta\approx13.6$ for the slightly over-doped sample, $x=0.17$,
and it increases (decreases) as the doping increases (decreases)
 \cite{Kimura}.
Thus, our calculation results are consistent with 
those of experiments.

According to the present numerical study,
the second term of eq.~(\ref{eqn:MC-lowT2}) 
is approximately proportional to the first term 
which is proportional to $\xi_{\rm AF}^{2x} \cdot \g_{\rm cold}^{-3}$.
Moreover, 
${\mit\Delta}\s_{xx}^{b}$ in eq.~(\ref{eqn:MC-lowT})
also makes an important contribution to the MR in LSCO,
which is proportional to $\xi_{\rm AF}^{2x}$.
In Fig.~\ref{fig:MKR-LSCO}, $\zeta$ in LSCO 
shows weak $T$-dependence, which mainly comes from 
the second term of eq.~(\ref{eqn:MC-lowT2}) 
 \cite{Comment2}.
However, its $T$-dependence is much smaller than that of 
$\zeta_{\rm RTA}$ or $\zeta_{\rm MC{\mbox{-}}RTA}$,
as shown in Fig.~\ref{fig:MKR-LSCO}.

In summary,
we obtain the following relations by taking the VC's into account
in the nearly AF Fermi liquid:
\begin{eqnarray}
{\mit\Delta}\s_{xy} \ &\propto& \ \xi_{\rm AF}^x \cdot \g_{\rm cold}^{-2},
 \nonumber \\
{\mit\Delta}\s_{xx} \ &\propto& \ \xi_{\rm AF}^{2x} \cdot \g_{\rm cold}^{-3},
 \label{eqn:scaling}
\end{eqnarray}
and $x\sim2$.
Because $\s_{xx}\propto \xi_{\rm AF}^0 \cdot \g_{\rm cold}^{-1}$,
we obtain the relations
${\mit\Delta}\rho/\rho\propto \xi_{\rm AF}^{2x} \cdot \g_{\rm cold}^{-2}$
and 
$\zeta \propto \xi_{\rm AF}^0 \cdot \g_{\rm cold}^{0} \approx $const.,
which explain why the "modified Kohler's rule" is 
satisfied in high-$T_{\rm c}$ cuprates in general
for a wider range of temperatures.

\begin{figure}
\vspace{10mm}
\begin{center}
\epsfig{file=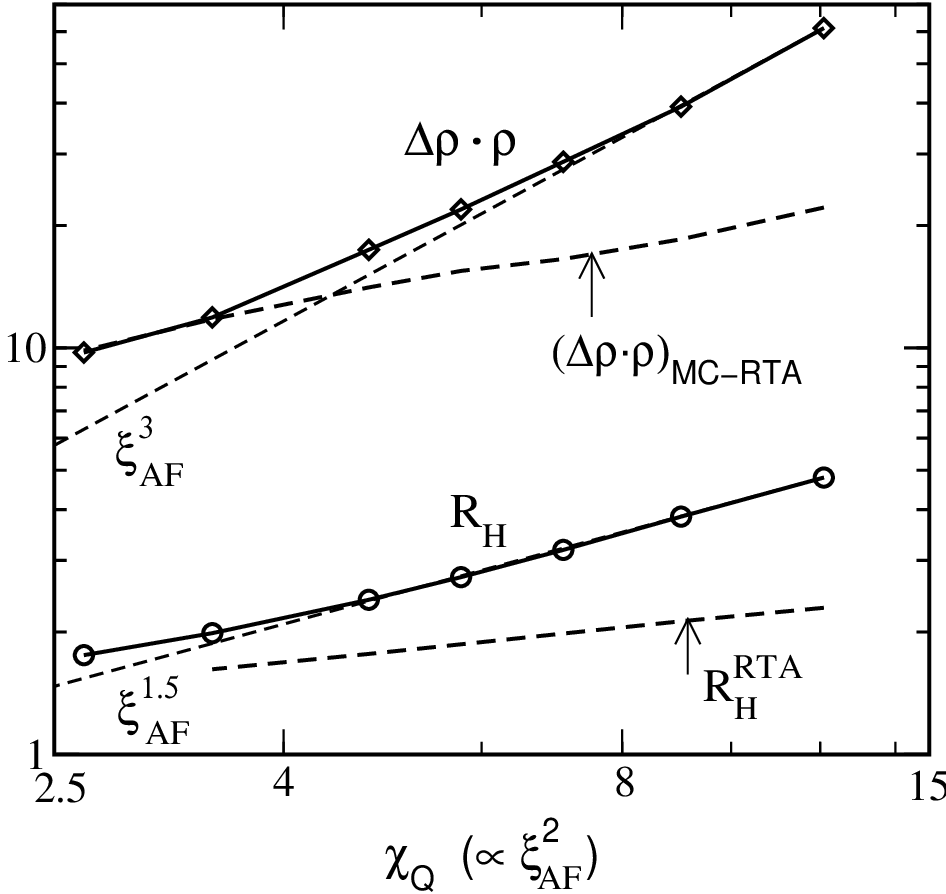,width=7cm}
\end{center}
\vspace{10mm}
\caption{}
{\small 
Scaling behaviors for YBCO ($n=0.9$);
${\mit\Delta}\rho\cdot\rho \propto R_{\rm H}^2 
\propto \xi_{\rm AF}^{2x}$,
which means $x\approx1.5$.
This plot is for 
$T=0.02\sim0.08$ ($\chi_Q= 12\sim2.6$).
We stress that the RTA fails to reproduce the 
scaling behavior.
}
  \label{fig:scaling-YBCO}
\end{figure}

Figure~\ref{fig:scaling-YBCO} shows that 
the obtained scaling behavior for YBCO,
$\Delta\rho\cdot\rho \propto R_{\rm H}^2 \propto \xi^{2x}$
and $x\approx 1.5$,
which is consistent with eq.~(\ref{eqn:scaling}) well.
This obtained result
directly means that ${\mit\Delta}\s_{xx} \propto T^{-4.5}$
or ${\mit\Delta}\rho/\rho \propto T^{-3.5}$
because the relations
$\chi_Q(0) \propto \xi_{\rm AF}^2 \propto T^{-1}$ 
and $\g_{\rm cold}\propto T^{-1}$ 
hold well for $T \ge 0.02$
in the present FLEX approximation.
This result is consistent with that of the experiment
on 90~K-YBCO
 \cite{Kimura}.
We note that in the present study
$R_{\rm H}(T=0.02)/R_{\rm H}(T=0.2)$ is about 3.8
for YBCO ($n=0.9$),
and about 4 for LSCO ($n=0.85$).

In this letter,
we mainly focus on systems around the optimum doping weight
above the pseudo spin-gap temperature ($T^\ast$),
where the modified Kohler's rule holds well experimentally.
In our numerical study, 
$\xi_{\rm AF}\simle 3a$ for $n\le0.9$ in YBCO 
($a$ being the lattice spacing),
which is consistent with experiments
 \cite{Kontani}.
Moreover, $\g_{\rm hot}/\g_{\rm cold}$ is suppressed to $\sim 3$
because of the feedback effect in the FLEX approximation.
It is an important theoretical (and experimental)
future problem to analyze the MC in a heavily under-doped region 
where $\xi_{\rm AF}\sim10$ will be realized.
Finally, we discuss the MR in the pseudo spin-gap region.
Experimentally, $\Delta\rho/\rho$ starts to increase
much faster than $T^{-4}$ below $T^\ast$.
This behavior can be explained if we assume that 
$\gamma_{\rm hot}/\gamma_{\rm cold} \gg 1$ is realized 
below $T^\ast$ because the second term of 
eq.~(\ref{eqn:MC-lowT2}) becomes much larger.
This assumption is supported by a recent study
of the pseudo spin-gap phenomena in terms of the 
strong superconducting fluctuations below $T^\ast$
 \cite{Yamase}.

In summary,
we studied the anomalous behaviors of the
MR in high-$T_{\rm c}$ cuprates
in terms of the Fermi liquid theory.
To go beyond the RTA,
the role of the VC's is analyzed by the conserving approximation
based on the exact expression for ${\mit\Delta}\rho/\rho$.
By virtue of the VC's,
the modified Kohler's rule is satisfactorily reproduced
in the presence of AF fluctuations.
The obtained value of $\zeta=\Delta\rho\cdot\rho/R_{\rm H}^2$
is consistent with experimental value both for YBCO and LSCO.
In conclusion, the VC's are essential for
reproducing the {\it seemingly} non-Fermi liquid behaviors of
both $R_{\rm H}$ and ${\mit\Delta}\rho/\rho$ in high-$T_{\rm c}$
on an equal footing.

The author is grateful to D. Vollhardt, K. Yamada and Y. Ando
for valuable comments and discussions.



\end{multicols}


\begin{thebibliography}{99}
\bibitem{Comment}
In contrast to ref.~\cite{Kontani},
here we put $|t'|>|t''|$ both for YBCO and LSCO
because it would be more realistic.
The results of the present work are also valid for
parameters $(t,t',t'';U)$ which are used in ref.~\cite{Kontani}.
\end{thebibliography}
\end{document}